\documentclass[a4paper]{spie}  
\usepackage[]{graphicx}
\usepackage{amstext} 
\usepackage{amssymb} 
\usepackage{amsmath}

\def\arcsec{\hbox{$^{\prime\prime}$}}

\title{First results from fringe tracking with the PRIMA fringe sensor unit} 

\author{J.~Sahlmann\supit{a,b,c}, R.~Abuter\supit{b}, S.~M\'enardi\supit{b}, C.~Schmid\supit{b}, N.~Di~Lieto\supit{b}, F.~Delplancke\supit{b}, R.~Frahm\supit{b}, N.~Gomes\supit{b}, P.~Haguenauer\supit{c}, S.~L\'ev\^eque\supit{b}, S.~Morel\supit{c}, A.~M\"uller\supit{b,d}, T.~Phan~Duc\supit{b}, N.~Schuhler\supit{c}, G.~van~Belle\supit{b}
\skiplinehalf 
\supit{a}Observatoire de Gen\`eve, Universit\'e de Gen\`eve, 51 Chemin Des Maillettes, 1290 Sauverny, Switzerland; \\
\supit{b}European Southern Observatory, Karl-Schwarzschild-Str. 2, 85748 Garching, Germany; \\
\supit{c}European Southern Observatory, Alonso de C\'ordova 3107, Vitacura-Santiago, Chile;\\
\supit{d}Max-Planck-Institut f\"{u}r Astronomie, K\"{o}nigstuhl 17, D-69117 Heidelberg, Germany
}
\authorinfo{Further author information: (Send correspondence to J.S.)\\J.S.: E-mail: johannes.sahlmann@unige.ch, Telephone: +41 223 79 2404}

\begin{document} 
  \maketitle 

\begin{abstract}
The fringe sensor unit (FSU) is the central element of the phase referenced imaging and micro-arcsecond astrometry
(PRIMA) dual-feed facility for the Very Large Telescope interferometer (VLTI). It has been installed at the Paranal observatory in August 2008 and is undergoing commissioning and preparation for science operation. Commissioning observations began shortly after installation and first results include the demonstration of spatially encoded fringe sensing and the increase in VLTI limiting magnitude for fringe tracking. However, difficulties have been encountered because the FSU does not incorporate real-time photometric correction and its fringe encoding depends on polarisation. These factors affect the control signals, especially their linearity, and can disturb the tracking control loop. To account for this, additional calibration and characterisation efforts are required. We outline the instrument concept and give an overview of the commissioning results obtained so far. We describe the effects of photometric variations and beam-train polarisation on the instrument operation and propose possible solutions. Finally, we update on the current status in view of the start of astrometric science operation with PRIMA.
\end{abstract}

\section{INTRODUCTION}\label{sec:intro} 
The PRIMA facility\cite{Delplancke:2008xr, van-Belle:2008eu} represents a major expansion of the VLTI infrastructure at Paranal Observatory in Chile. It introduces dual-feed capability and is designed for observation of two stellar objects within the isoplanatic angle of typically 0.5 arc-minutes. It consists of star-separator modules at the telescopes (STS\cite{Nijenhuis:2008cy}), differential delay lines (DDL\cite{Pepe2008}), an internal laser metrology\cite{Schuhler2007}, and the fringe sensor unit (FSU\cite{Sahlmann:2009kx}) in the beam-combination laboratory. The scientific drivers for the development of PRIMA are narrow-angle astrometry\cite{Launhardt2008} and phase referenced imaging in conjunction with the MIDI and AMBER instruments. The general status of the PRIMA facility is presented in Ref.~\citenum{Schmid2010}. Here, we present the status and results for the FSU. 

The FSU operates in the infrared $K$-band. It consists of two identical fringe sensors (FSUA and FSUB) and uses polarisation-dependent spatial encoding to obtain 4 simultaneous ABCD signals with relative phase-shifts of about 90 degrees. The beams are combined before spatial filtering by single-mode fibres to allow the injection of the internal laser metrology. A cryogenic low-resolution spectrograph disperses the light on 5 spectral pixels in each ABCD quadrant. The white-light pixel is not a physical one, but is synthesised from the sum of the 5 spectral pixels. Simultaneous estimates of group delay (GD) and phase delay (PD) are obtained from the spectral pixels and the white-light pixels, respectively. The FSU does not incorporate real-time photometric monitoring, i.e. there is no real-time access to the intensity in each interferometer arm or in the individual quadrants. For a detailed description of the instrument, the operation principle, and the observation setup, we refer the reader to Ref.~\citenum{Sahlmann:2009kx}.

After concluding the laboratory tests in Europe\cite{Sahlmann2008a}, the FSU was installed at Paranal observatory in July and August 2008. First fringes and fringe tracking on a stellar object were obtained one month later. The FSU commissioning activities since then include the establishment of the observation range in terms of stellar magnitude and atmospheric conditions, the detailed assessment of the fringe tracking performance, and the automatisation of fringe detection and tracking. Caused by the un-availability of PRIMA subsystems, the commissioning of the PRIMA facility started later and first functional tests of the FSU together with other PRIMA subsystems were performed. Simultaneous fringe tracking on two stellar objects has not been achieved at the time of writing\cite{Schmid2010}, but is expected to be achieved within a few months.  

One important result obtained from on-sky fringe tracking is that the group delay and phase delay may not be robust against the perturbations that occur during observation. The FSU signals can be highly non-linear, which disturbs the fringe tracking loop and eventually affect the quality of planned scientific programs. Therefore, most of this contribution is concerned with the problem of non-linearity.  

\begin{figure}[t]
\begin{center} 
\includegraphics[width= \linewidth, bb= -153   246   796   566]{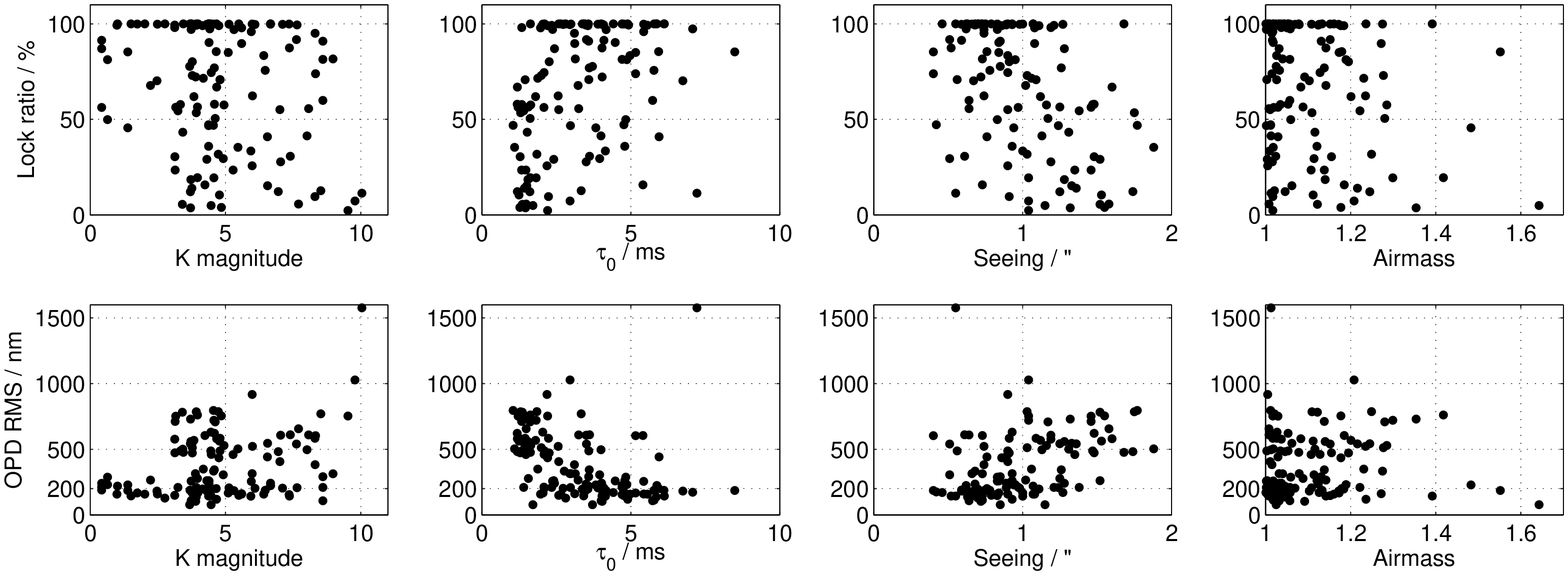} \end{center} 
\caption{Lock ratio over one minute (\textit{top row}) and residual OPD over one second (\textit{bottom row}) as function of \textit{K}-band magnitude, coherence time, seeing, and airmass. The Figure is from Ref.~\citenum{Sahlmann:2009kx}.}
\label{fig:lr}
\end{figure}

\section{FIRST RESULTS FROM SINGLE-FEED FRINGE TRACKING}
Results of FSU single-feed fringe tracking have already been published\cite{Sahlmann:2009kx}. The main conclusions were:
\begin{enumerate}
  \item Using the FSU, the limiting magnitude for fringe tracking with Auxiliary Telescopes (AT) at VLTI is improved to $\sim m_K = 7$, which is about 2 magnitudes fainter than the FINITO\cite{Bouquin2008} limit of $m_H = 5$\footnote{FINITO offers 3-telescope fringe tracking, whereas the FSU operates with 2 telescopes.}(Fig.~\ref{fig:lr}). Fringe tracking on $m_K = 9$ was demonstrated. In a first test, the improved sensitivity was also shown using the 8-m Unit Telescopes when fringes were detected on a $m_K = 11.7$~mag star.
  \item The FSU operation range in terms of atmospheric conditions agrees with the experience of the first VLTI fringe tracking sensor FINITO operating in $H$-band and is approximately defined by coherence time $\tau_0 > 2.5$~ms, seeing $<1.2\arcsec$, and airmass $< 1.5$. Also the FSU fringe tracking precision in terms of residual OPD is comparable to FINITO.
  \item At present, the accuracy of FSU fringe tracking is problematic. It can not always be guaranteed that the FSU is tracking the central fringe (Fig. \ref{fig:fringing}). This represent a potential problem for future scientific observations, because it can introduce systematic biases that can not be averaged out.  
\end{enumerate}
Additionally, functional tests of fringe tracking with the DDL-actuators and using one feed of the STS were performed\cite{Schmid2010}. FSU fringe tracking and simultaneous on-axis observation with the MIDI instrument was demonstrated and shows significant improvements concerning the gain in sensitivity and data quality of MIDI\cite{Mueller2010}.

\section{CALIBRATING THE FSU}
\begin{figure}
\begin{center} 
\includegraphics[width= 0.49\linewidth]{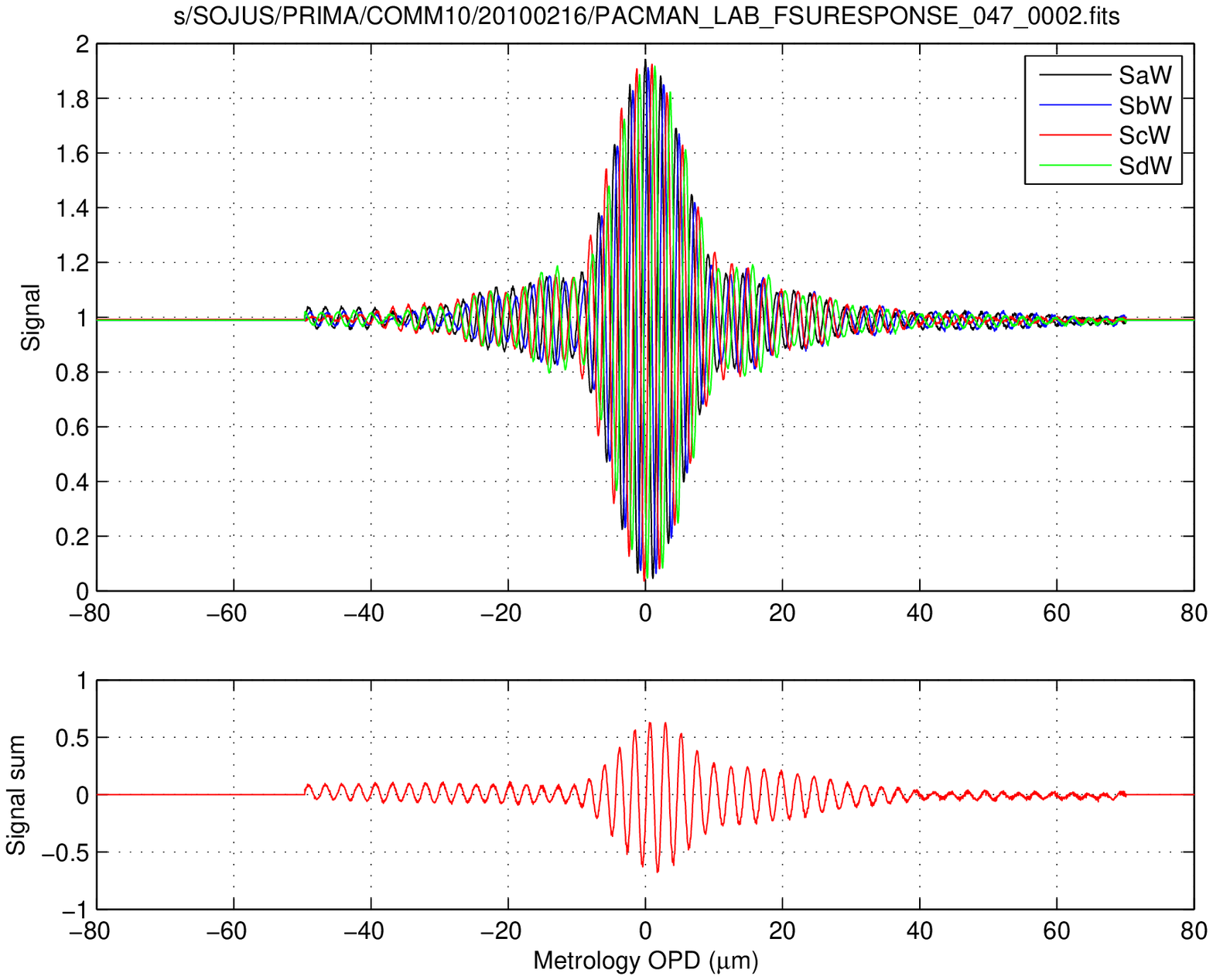}
\includegraphics[width= 0.49\linewidth]{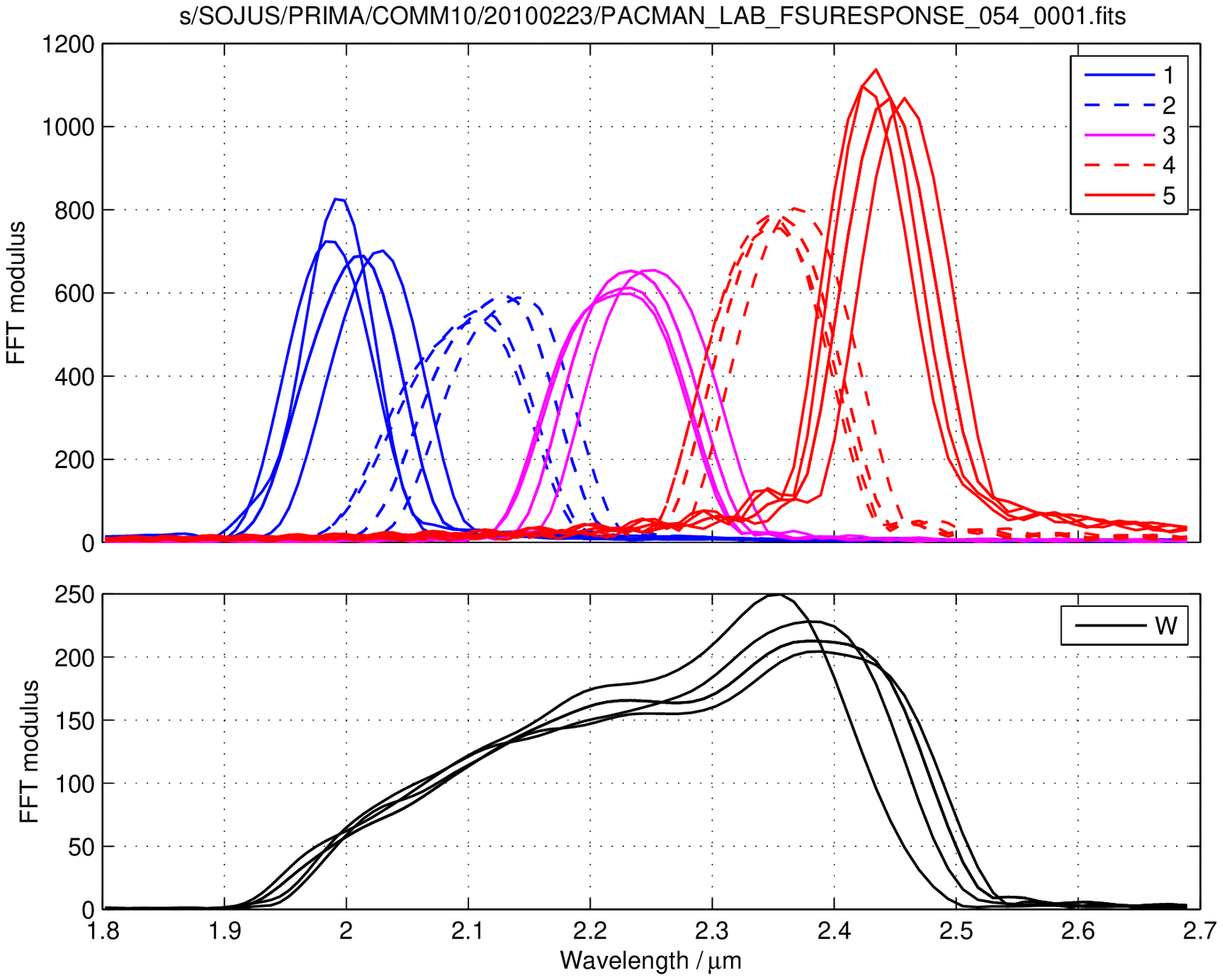}
 \end{center} 
\caption{Snapshots of a typical laboratory calibration of FSUB. \emph{Left}: White light fringes obtained from the calibration scan in OPD. The bottom panel shows the appearance of fringes in the sum of ABCD, which ideally should be constant across the fringes. \emph{Right}: The modulus of the corresponding Fourier transform. The top panel shows the 20 physical pixels and the wavelength dispersion across ABCD within the same spectral channel. The bottom panel shows the synthetic white light pixel and exhibits a slope corresponding to the spectral energy distribution of the calibration black-body light source at 700 C$^\circ$.}
\label{fig:fft}
\end{figure}

The calibration principle yielding the photometric, wavelength, and phase-shift parameters in the laboratory that are required for the real-time algorithms has been presented previously\cite{Sahlmann2008b,Sahlmann:2009kx}, and we discuss further details here. The calibration is based on Fourier transform spectroscopy. We verified by simulation that it is accurate to the level of 0.1~nm in wavelength and to $<1\deg$ in phase-shift. However there are several factors that limit the FSU performance even when wavelengths and phase-shifts are accurately known:
\begin{enumerate}
  \item The FSU does not incorporate real-time photometric monitoring of each telescope beam\footnote{Ideally, the sum of ABCD bins could be used as real-time photometric estimator with the limitation that it does not give access to the relative intensity ratio between the two input beams. In practice, this estimator is not robust because of losses in the opto-mechanical system, primarily in the beam-combiner, that create a delay-dependent modulation of the sum of ABCD (see Fig. \ref{fig:fft} bottom left).}. Photometric calibration in the laboratory and on-sky is performed before the observation. Any deviation of the actual intensities from the calibrated ones, which is different for A, B, C, and D channels, degrades the quality of FSU measurements. 
  \item The current algorithms do not take into account the wavelength dispersion across ABCD within each spectral band (cf. Fig. \ref{fig:fft}).  
\end{enumerate}  
\subsection{Figure of merit: the linearity of FSU-measured delays} 
Since many parameters influence the calibration process, a simple observable has to be defined to classify a given calibration as \emph{good} or \emph{poor}. A suitable parameter for this purpose are the linearities of optical path difference (OPD, derived from phase delay) and group delay measured by the FSU. The validation of a calibration therefore consists of computing the OPD and GD estimates on the calibration data, i.e. the OPD scan data, themselves. The reference delay $OPD_\mathrm{true}$ is measured by the internal laser metrology. We measure non-linearity in terms of the quantity 
\begin{equation}
q_{OPD} = \frac{\partial OPD_\mathrm{measured}}{\partial OPD_\mathrm{true}} - 1 \hspace{1cm} q_{GD} = \frac{\partial GD_\mathrm{measured}}{\partial OPD_\mathrm{true}} - 1,
\end{equation}
thus in the ideal case of a linear signal $q=0$. To remove high-frequency noise, the measured delays are low-pass filtered with a moving average of length $\lambda/10$. 

\begin{figure}[t]
\begin{center} 
\includegraphics[width= 0.49\linewidth]{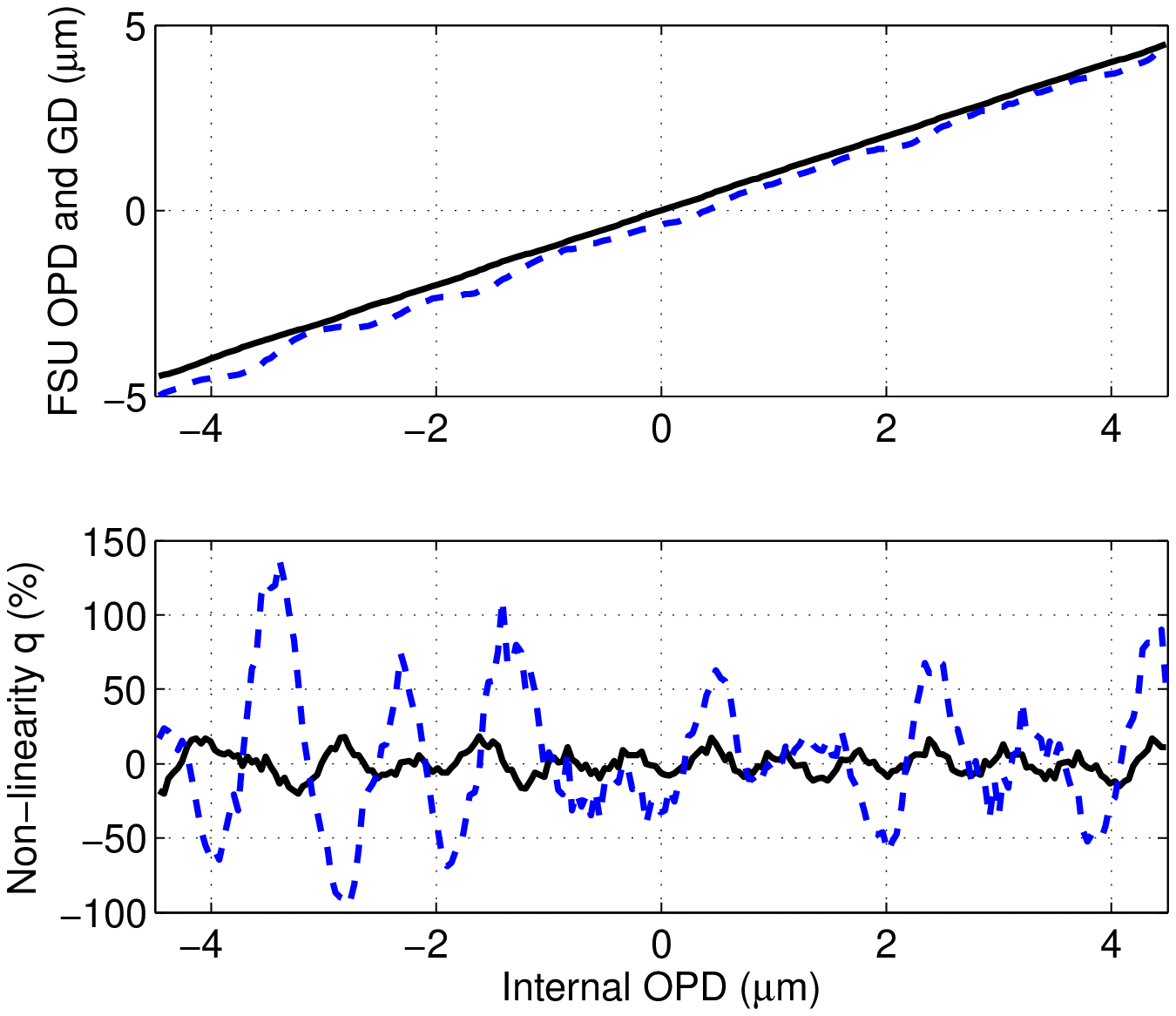}
\includegraphics[width= 0.49\linewidth]{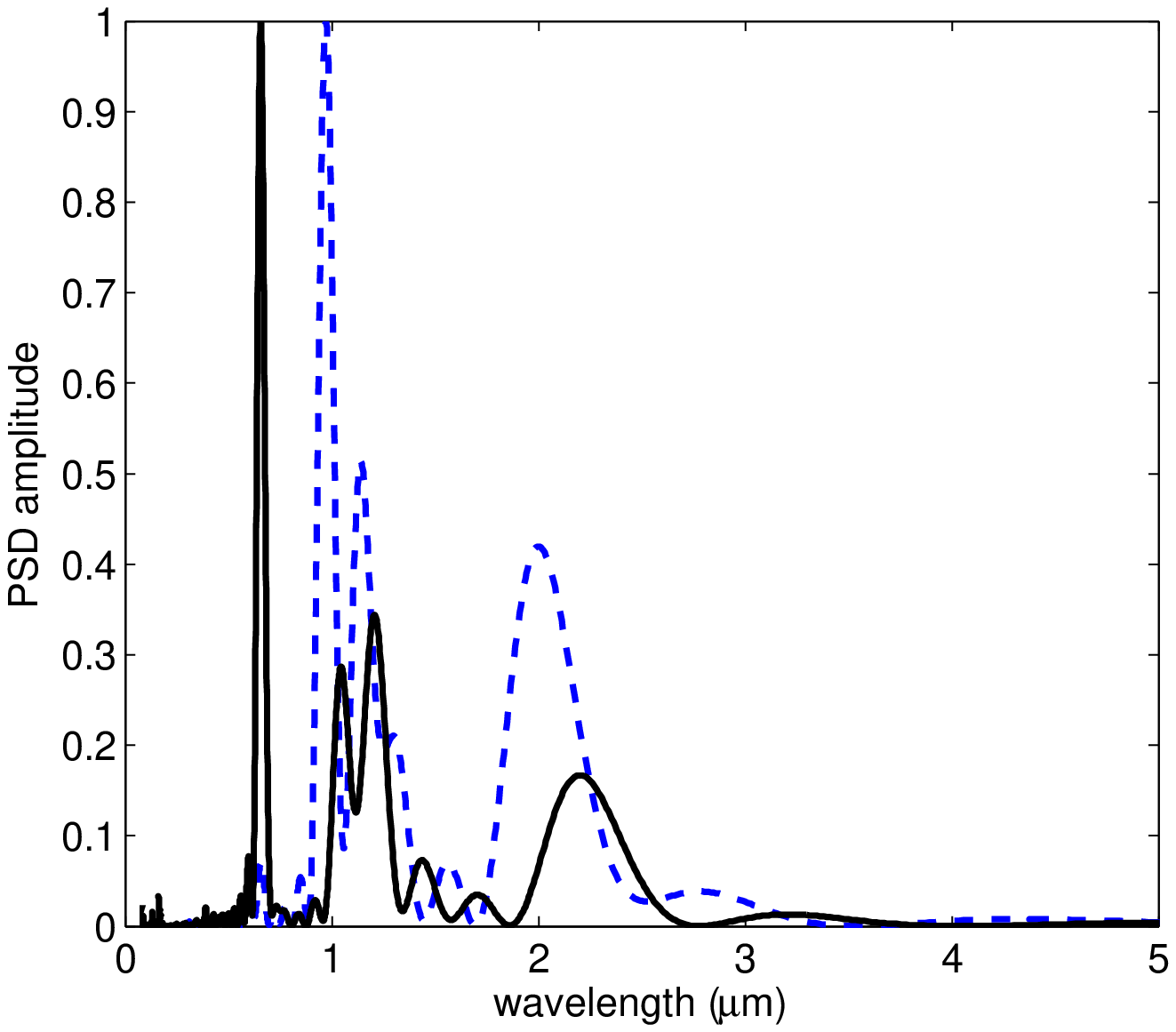}
 \end{center} 
\caption{OPD (black solid lines) and GD (blue dashed lines) linearity for a typical laboratory calibration of FSUB. \emph{Left}: The top panel shows the OPD and GD obtained within the central 9 $\mu$m of the calibration scan. The bottom panel shows the corresponding non-linearity parameter $q$. \emph{Right}: The normalised spectra of the non-linearities $q_\mathrm{OPD}$ and $q_\mathrm{GD}$}
\label{fig:linearities}
\end{figure}

Typical results are shown in Fig. \ref{fig:linearities}. The OPD non-linearity with peak and root-mean-square value of $\pm20$\% and 8~\%, respectively, is generally lower than the GD non-linearity with peak and root-mean-square value of $\pm120$\% and 45~\%, respectively. Also the OPD non-linearity amplitude remains approximately constant over the 9 $\mu$m range, whereas the GD non-linearity amplitude increases with the distance to the central fringe. Periodicity or \emph{cyclic errors} are a clear feature of these non-linearities and can be seen in the power-spectral density. The analysis of these spectra, revealing their respective cyclic wavelength $\lambda_\mathrm{cyclic}$ and the wavelength ratio $\rho_\mathrm{cyclic}$ compared to the effective wavelength in the white-light channel $\lambda_\mathrm{eff,0}$, is summarised in Table \ref{tab:fonts}. 

\begin{table}[h]
\caption{Periodicity in OPD and GD non-linearity evaluated on a calibration scan.} 
\label{tab:fonts}
\begin{center}       
\begin{tabular}{|r|r | r | r | r |} 
\hline
\rule[-1ex]{0pt}{3.5ex}  Delay type & Rel. PSD amplitude & $\rho_\mathrm{cyclic}$ & $\lambda_\mathrm{cyclic}$ &$\lambda_\mathrm{eff,0}$\\
\hline
\rule[-1ex]{0pt}{3.5ex}  - & (\%) & - & ($\mu$m) & ($\mu$m)\\
\hline
\hline
\rule[-1ex]{0pt}{3.5ex}  OPD & 100 & 3.41 & 0.655 &2.236\\
\hline
\rule[-1ex]{0pt}{3.5ex}  OPD & 34 & 1.85 & 1.206 &2.236\\
\hline
\rule[-1ex]{0pt}{3.5ex}  OPD & 26 & 2.11 & 1.062 &2.236\\
\hline
\rule[-1ex]{0pt}{3.5ex}  OPD & 17 & 1.02 & 2.201 &2.236\\
\hline
\hline
\rule[-1ex]{0pt}{3.5ex}  GD & 100 & 2.30 & 0.971 &2.236\\
\hline
\rule[-1ex]{0pt}{3.5ex}  GD & 42 & 1.12 & 2.000 &2.236\\
\hline
\rule[-1ex]{0pt}{3.5ex}  GD & 30 & 1.97 & 1.137 &2.236\\
\hline
\rule[-1ex]{0pt}{3.5ex}  GD & 7 & 3.39 & 0.658 &2.236\\
\hline
\end{tabular}\end{center}\end{table} 

\section{SOURCES OF SIGNAL NON-LINEARITY}
As shown in the previous section, periodic non-linearities occur in the phase and group delay estimates of the FSU. The appearance of these \emph{cyclic errors} is a known problem in length measurements using interferometry\cite{Quenelle1983, Wu:1996la}. We list here the causes of non-linearity that apply to the case of the FSU. For analytical considerations we employ a simple model, in which the white light ABCD quadrature signals of the FSU are described as phase-shifted sine-waves\footnote{This is of course an idealised model. In reality, the phase-shifts can be far from 90$^\circ$. The algorithm employed in operation takes these non-ideal phase-shifts into account\cite{Sahlmann:2009kx}.}:
\begin{equation} 
A = \cos \phi \hspace{1cm} B = \sin \phi  \hspace{1cm} C = -\cos\phi \hspace{1cm} D = -\sin \phi
\end{equation}
\begin{equation}\label{eq:AB}
\tan \phi = \frac{B}{A}
\end{equation}
For the purpose of this derivation the phase $\phi$ is derived from the A and B signals only (Eq. \ref{eq:AB}). However, the results remain valid when using the nominal algorithm (Eq. \ref{eq:nominal}). In the definition of error sources, we use the term \emph{differential} to designate effects that are different in A, B, C, and D. If instead the effect is common to all quadrants, we use the term \emph{common-mode}.
\begin{enumerate}
  \item \textbf{Differential photometric error:} Let us assume the intensity of channel B has not been properly calibrated and exhibits a residual photometric offset $\Delta$ with respect to A. If we are not aware of the mis-calibration, we measure a false phase $\phi'$:
  \begin{equation}
\tan \phi' = \frac{B + \Delta}{A} = \frac{\sin (\phi) + \Delta}{\cos \phi} = \frac{\Delta + \tan \phi \cos \phi}{\cos \phi } = \frac{\Delta}{\cos \phi} + \tan \phi.
\end{equation}
The measured phase $\phi'$ is thus equal to the true phase $\phi$ modulated with an additional term of same periodicity. The non-linearity amplitude increases with increasing photometric error $\Delta$. It is trivial to verify that  $\tan \phi' \rightarrow \tan \phi$ for $\Delta \rightarrow 0$. 
  \item \textbf{Differential phase-shift error:} If the phase-shift of one quadrant is not correctly derived by the calibration and is wrong by an amount $\delta$, the measured phase $\phi'$ is:
  \begin{equation}
\tan \phi' = \frac{B'}{A} = \frac{\sin (\phi + \delta)}{\cos \phi} = \frac{2}{\cos(2\phi) +1} \cos \phi \sin(\phi+\delta) = \frac{\sin(2\phi+\delta) + \sin \delta}{\cos(2\phi) +1} 
\end{equation}
We thus measure a false phase, whose value depends on $2\phi$. Therefore, the resulting non-linearity is found at twice the frequency. It is easily verified that $\tan \phi' \rightarrow \tan \phi$ for $\delta \rightarrow 0$.
  \item \textbf{Common-mode wavelength error:} If all wavelengths are falsely derived and exhibit a common bias, the effective wavelength $\lambda_\mathrm{eff,0}$ will have the same bias. Because the calibration uses this effective wavelength to derive the relative phase-shifts between A, B, C, and D, the consequence of such a bias is that the estimated phase-shifts are wrong by a common factor, appearing as an increasing error for B/A, C/A and D/A phase shifts. These differential phase-shift errors cause a non-linear phase signal as described in point~2.
    \item \textbf{Common-mode photometric error:} These errors can occur if a change in transmission or alignment of the instrument is not calibrated. In the ideal case and when using the nominal algorithm\footnote{In the real system, the ABCD signals have been normalised before performing this calculation.} 
  \begin{equation}\label{eq:nominal}
\tan \phi = \frac{B-D}{A-C}
\end{equation} 
a common-mode photometric error has no effect on the phase measurement. In reality, however, any photometric variation is unlikely to be common-mode, because (a) the FSU-internal transmissions of ABCD are different for beam 1 and beam 2 and an un-calibrated common-mode photometric variation (e.g.  caused by the atmospheric transmission) results in a differential photometric error; (b) the photometric variations originating in uncorrected tip-tilt errors are always differential, because of residual misalignment between the four ABCD single-mode fibres\cite{Sahlmann:2009kx}. The resulting phase non-linearity is described in point 1.
	\item \textbf{Differential wavelength error:} For the FSU, the wavelengths across the quadrants ABCD are not identical, but differ by a few percent (Fig. \ref{fig:fft}). The applied algorithm does not take this dispersion into account, but uses instead the average of the individually measured ABCD effective wavelengths. 	Because the calibration uses this effective wavelength to derive the relative phase-shifts between A, B, C, and D, these phase-shifts are certainly erroneous\footnote{In other words: with differential wavelengths, the phase-shifts become linearly dependent on OPD, which is not supported by the current phase algorithm.}, which leads to the situation of point 2. 
\end{enumerate}
We have now found a way to relate the error sources and the effect on the non-linearity of the measured phase (or equivalently OPD). In terms of the cyclic wavelength ratio, two scenarios were deduced: \begin{description}
  \item[$\rho_\mathrm{cyclic} = 1$]: The cyclic phase error has the same periodicity as the carrying wavelength $\lambda_\mathrm{eff,0}$ and is caused by differential photometric errors. 
  \item[$\rho_\mathrm{cyclic} = 2$]: The cyclic phase error has twice the frequency of the carrying wavelength $\lambda_\mathrm{eff,0}$ and is caused by differential phase-shift errors. 
\end{description}  
\begin{figure}
\begin{center} 
\includegraphics[width= 0.49\linewidth, trim = 0cm 0cm 0cm 0.5cm, clip=true]{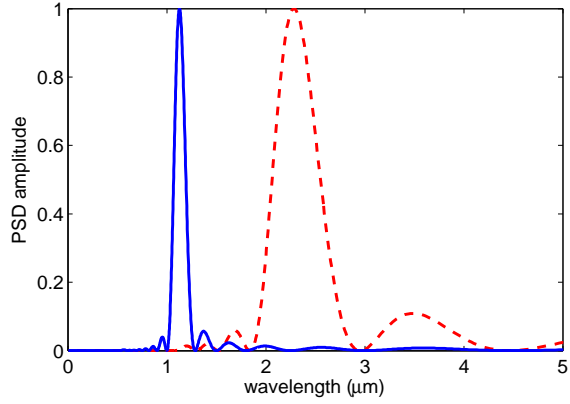}
 \end{center} 
\caption{Simulated OPD non-linearity spectrum caused by differential photometric errors (dashed red line, $\rho_\mathrm{cyclic} = 0.984$) and by differential phase-shift errors (blue solid line, $\rho_\mathrm{cyclic} = 2.004$).  }
\label{fig:simuq}
\end{figure}
These characteristics were reproduced in simulation and are shown in Fig \ref{fig:simuq}. Although not demonstrated here, we can expect that the same effects apply to the GD non-linearity because it is based on the same measurement principles. In fact, the GD non-linearity spectrum exhibits a similar pattern, as is shown in Fig. \ref{fig:linearities}. Except for the highest-frequency peak of OPD non-linarity at $0.65 \mu$m, the GD spectrum is very similar to the OPD spectrum, though slightly shifted towards smaller wavelengths. This may indicate that the overall effective wavelength for GD computation is smaller than for OPD computation. The peaks close to $2 \mu$m can be explained by differential photometric errors. The peak right above $1 \mu$m is caused by differential phase-shift errors, although the reason for the appearance of a double peak is not reproduced by simulation. \\
At this point, there is no proven explanation for the OPD non-linearity peak at $0.655\, \mu$m. However, it is likely that it is related to a cyclic error of the PRIMA laser metrology, which operates at $\lambda=1.319 \,\mu$m ($=2\times 0.6595 \,\mu$m). A probable cause for this cyclic error is polarisation leakage due to non-perfect alignment. 
 
\subsection{Sources of signal non-linearity during on-sky fringe tracking}\label{sec:sourc}
During the observation of a stellar object, a large variety of perturbations can occur and disturb the GD/OPD linearity, hence the fringe tracking loop: 

Currently, the FSU is calibrated in the laboratory only, thus the calibration parameters for wavelengths and phase-shifts derived in the laboratory are also applied for on-sky observation. Because of the differences in source spectral energy distribution (a $700^\circ$C black body in the laboratory and a stellar object during observation) and in the beam train (during observation, about a dozen additional reflections are required to feed the stellar light into the laboratory), we can expect that phase-shifts and wavelengths in the laboratory and on sky are different. This will introduce errors and non-linearities in the delay computation. A solution to this problem is to perform the FSU calibration on-sky. This is planned, once the dual-feed system is available. The temporal variability of the calibration parameters and the achievable precision on faint targets will have to be investigated. 

In addition, the FSU design does not incorporate dedicated real-time photometric monitoring. Therefore the photometric calibration has to be performed before (and possibly after) the observation. Photometric variability at timescales of the observation perturbs the real-time estimates. Examples of such photometric variability are (a) low-frequency changes in VLTI-transmission (b) high-frequency differential injection caused by tip-tilt jitter. If this is the limiting factor, the cadence of photometric calibration has to be increased.

All these effects occur simultaneously and are therefore difficult to distinguish. At this point, there is no robust way to measure the phase and group delay non-linearities on-sky, which could give insight in their causes through spectral analysis as in the previous section. 

\section{FRINGE TRACKING LOOP STABILITY}
\begin{figure}[b]
\begin{center} 
\includegraphics[width= 0.53\linewidth]{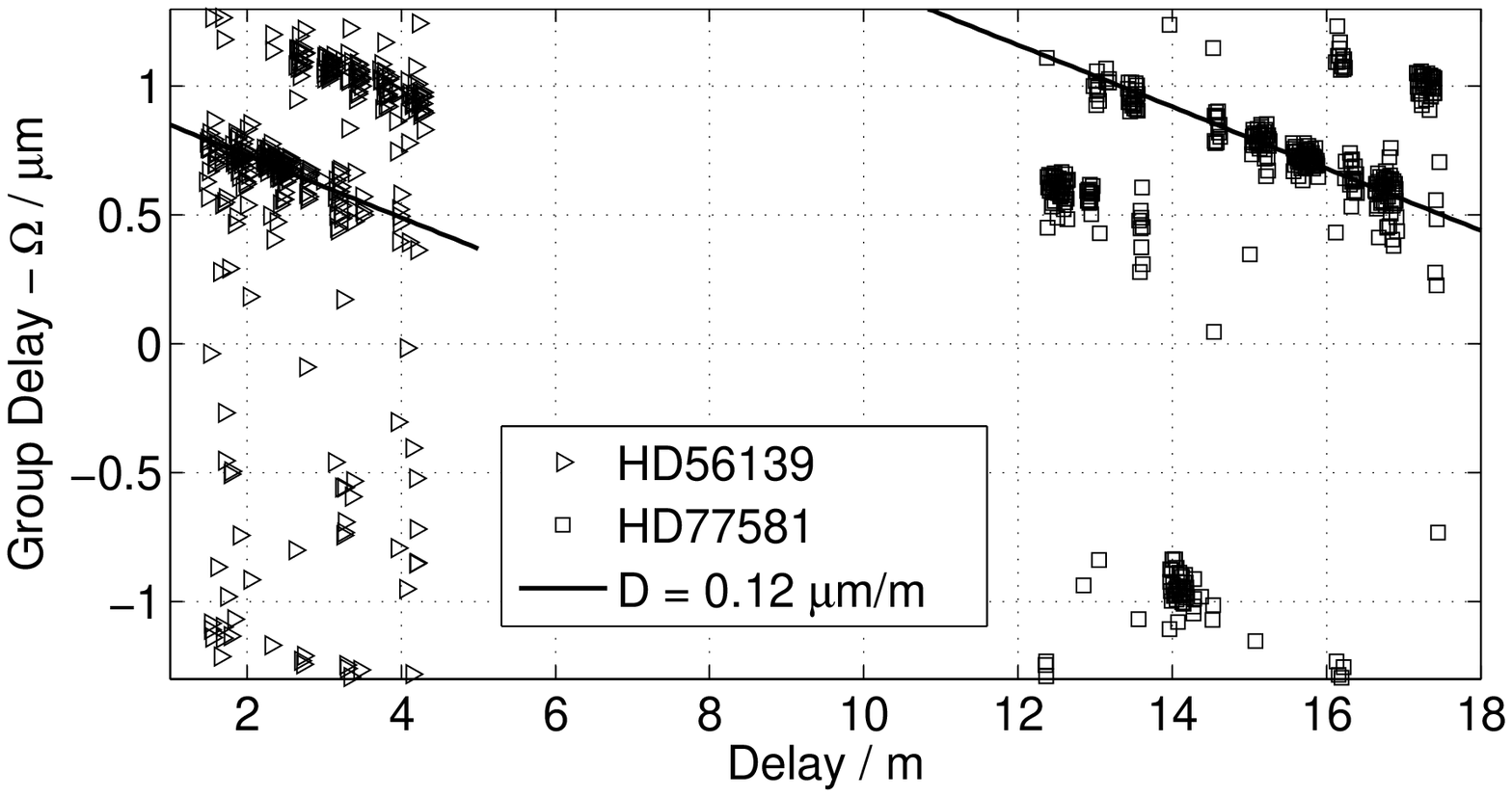}
\includegraphics[width= 0.46\linewidth]{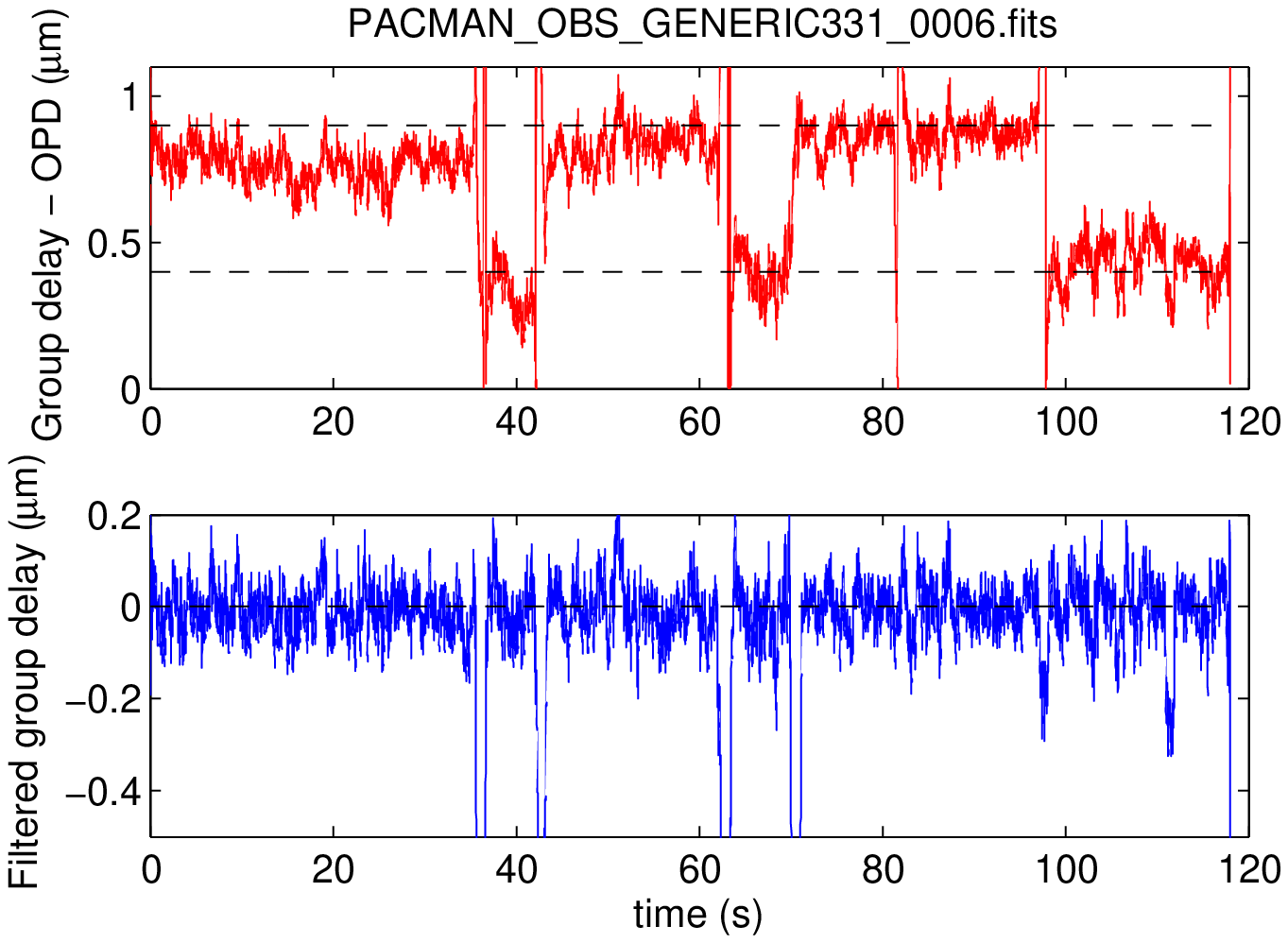}
 \end{center} 
\caption{\emph{Left}: Group delay minus FSU OPD as function of VLTI delay. Each point corresponds to 1~s of data (Figure from \cite{Sahlmann:2009kx}). \emph{Right}: The top panel shows GD - OPD during one fringe-tracking observation. Two levels separated by $\sim 0.5 \mu$m are seen. The bottom panel shows the GD measurement only, low-pass filtered by a moving average of 0.5 s. No significant deviation from the target value of 0 is apparent (the downwards spikes occur when the tracking loop is temporarily open).}
\label{fig:fringing}
\end{figure}

\begin{figure}
\begin{center} 
\includegraphics[width= 0.7\linewidth]{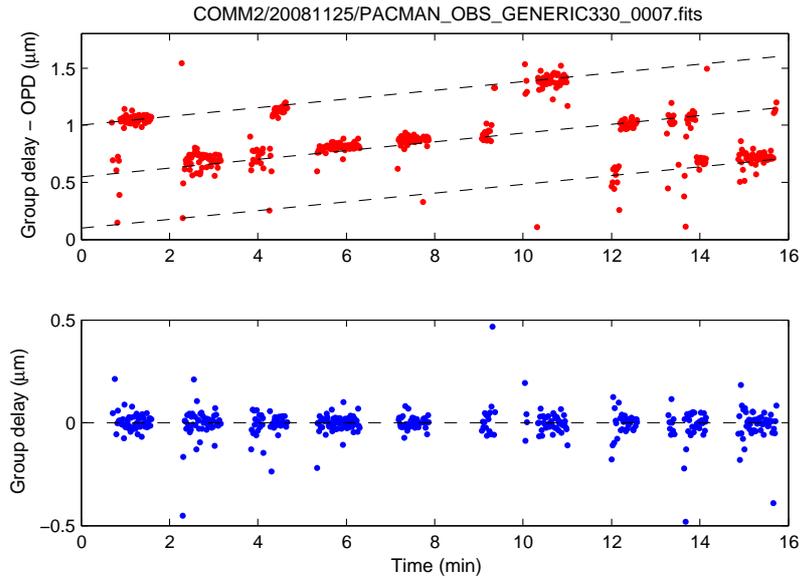}
 \end{center} 
\caption{The top panel shows GD - OPD during a long fringe-tracking observation of 15 minutes. Data is shown only if the fringe tracking loop was closed. Each point corresponds to 1~s of data. Three levels separated by $\sim 0.45 \mu$m are seen. The bottom panel shows the GD measurement only. No significant deviation from the target value of 0 is apparent.}
\label{fig:fringing2}
\end{figure}

There are many parameters that can disturb the FSU real-time estimates of OPD and GD. Since the FSU acts as sensor of the fringe tracking control loop, the quality of the OPD and GD measurements is essential for the stability of the control system. While Fig. \ref{fig:lr} is useful to evaluate the fringe tracking precision, it does not display the tracking accuracy, i.e. it can not be used to decide whether the central fringe is tracked. First concerns in this direction were raised based on the behaviour of the value of group delay minus phase delay during fringe tracking, which showed several distinct levels (see Figs. \ref{fig:fringing} and \ref{fig:fringing2})\cite{Sahlmann:2009kx}. The conclusion was that tracking within the 3 central fringes is ensured. If fringe-hops occur and are not taken into account in the data reduction of e.g. an astrometric observation, they introduce systematic biases, that can not be averaged out.

\begin{figure}[b]\begin{center} 
\includegraphics[width= 6cm, angle = -90, trim = 6cm 4cm 6cm 4cm, clip=true]{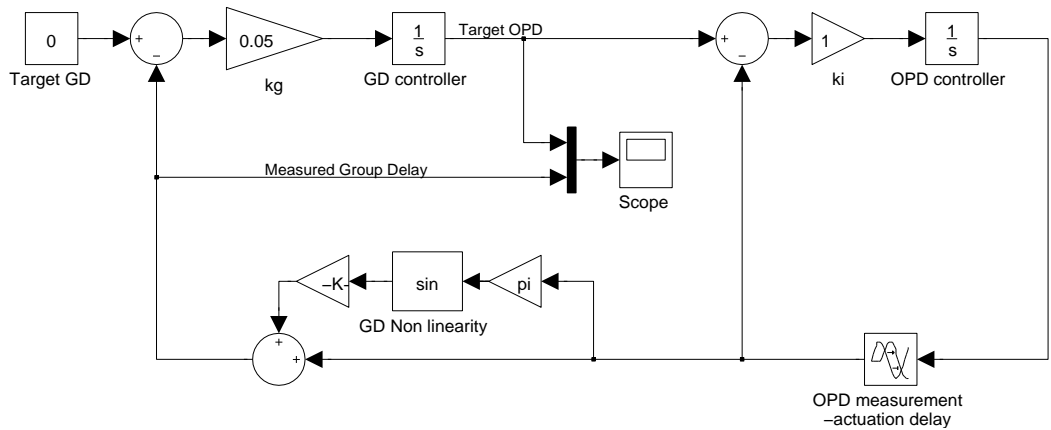} \end{center} \caption{\texttt{Simulink} model of the PRIMA fringe tracking control loop. The inner OPD loop is located on the right side. The outer GD loop including the non-linearity injection is located towards the left of the graph. Within this control system, the OPD is controlled to a variable target such that the group delay remains at its target value of 0.}\label{fig:loopmodel}
\end{figure}

We can observe several effects in Figs. \ref{fig:fringing} and \ref{fig:fringing2}: A linear trend of GD--OPD as function of total delay is the effect of atmospheric dispersion\cite{Colavita2004, Sahlmann:2009kx} and not discussed here. In addition, two or three levels are observed which are approximately evenly spaced by $\sim 0.5 \mu$m. Transitions between these levels usually occur when the control loop opens for short time and locks again (due to a strong perturbation or when the observation requires the opening). Meanwhile, the group delay is constantly kept close to zero by the fringe tracking loop. A possible interpretation\cite{Sahlmann:2009kx} is that the FSU estimate of group delay is non-linear and exhibits several zeroes or local extrema close to zero. The difference between measured GD and OPD would then depend on the position within the fringe packet because of the combination of GD and OPD non-linearities. This scenario is investigated in the following.
\begin{figure}\begin{center} 
\includegraphics[width= 0.49\linewidth]{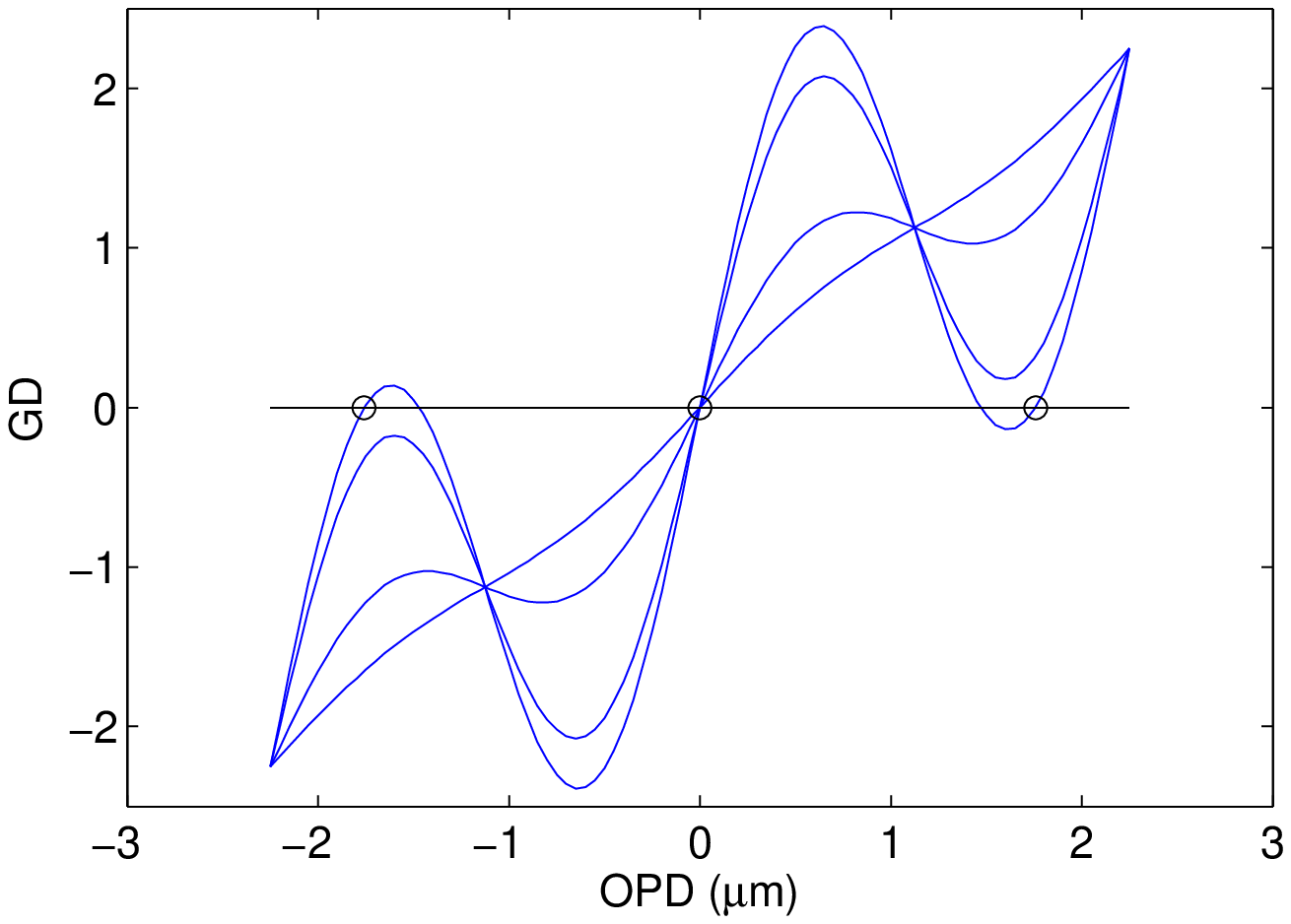}
\includegraphics[width= 0.49\linewidth]{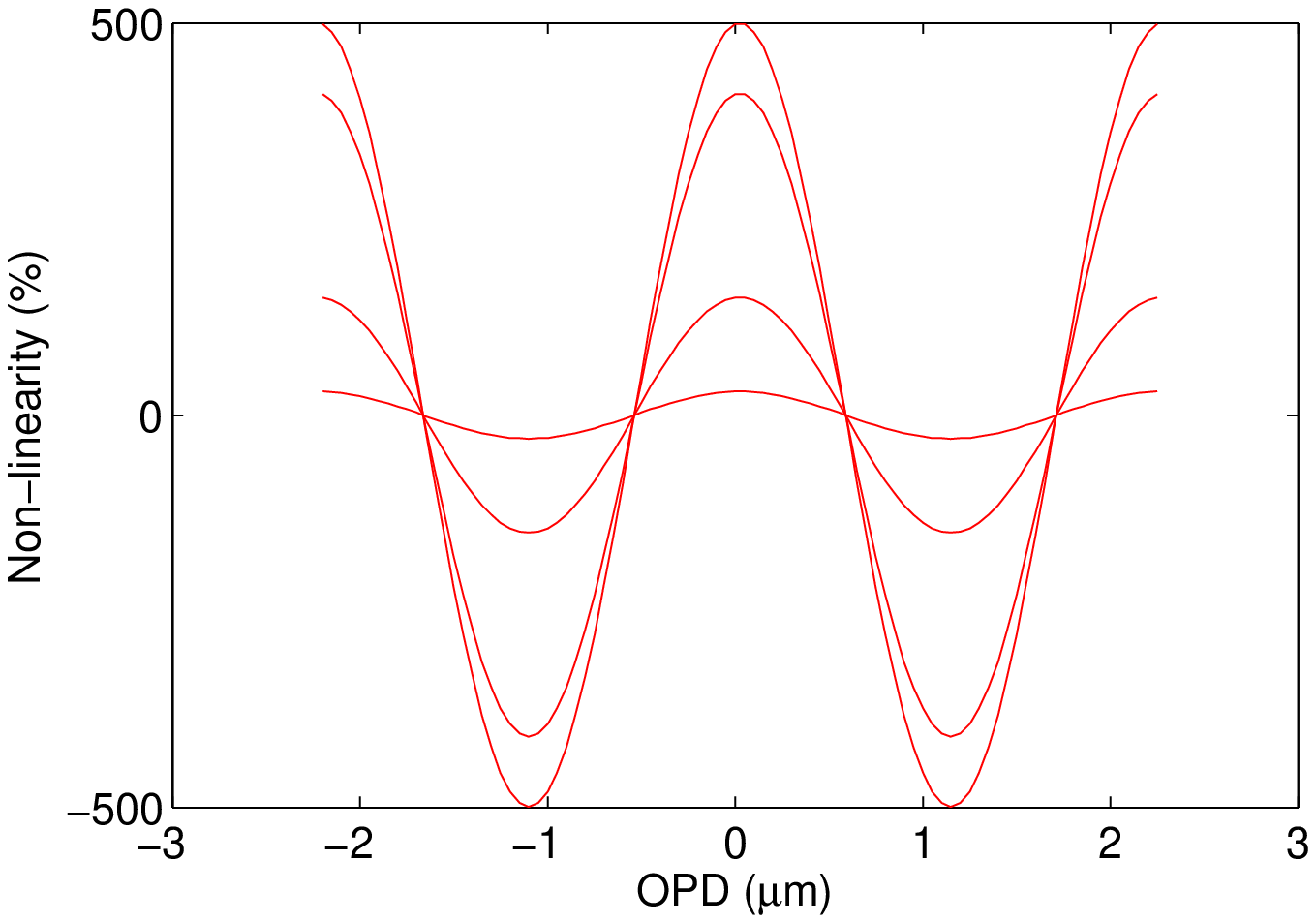} \end{center} 
\caption{Simulated group delay (left) and group delay non-linearity (right) for $\hat q =$ 30 \%, 100 \%, 410 \%, and 500 \% and $\nu =1$. Stable attractive points are indicated with black circles. For moderate non-linearity there is only one zero (and therefore one stable point) and multiple local maxima and minima. For $\hat q = 500$ \%, there are three stable points.}
\label{fig:GDsimu}
\end{figure}

\begin{figure}[t]\begin{center} 
\includegraphics[width= 0.8\linewidth]{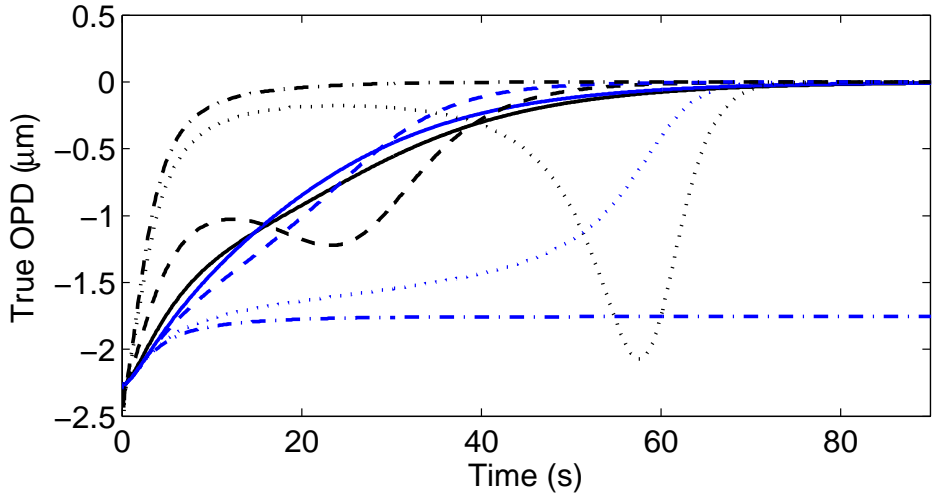} \end{center} \caption{Simulation of the loop behaviour for $\hat q = 30$\,\% (solid lines), $\hat q = 100$\,\% (dashed lines), $\hat q = 410$\,\% (dotted lines), and $\hat q = 500$\,\% (dash-dotted lines) with cyclic frequency $\nu=1$. The measured GD is shown in black and the target phase delay is drawn in blue. The initial conditions are set to start the simulation at minus one wavelength of OPD.}\label{fig:convergence}
\end{figure}
\subsection{SIMULATION OF THE FRINGE TRACKING CONTROL SYSTEM }
In the previous sections, we showed that the FSU real-time estimates of GD and PD show considerable non-linearity when evaluated in laboratory conditions \cite{Sahlmann2008b, Sahlmann:2009kx}. These non-linearities are expected to be larger during on-sky observation because of the uncertainty in the calibration parameters (cf. Sect. \ref{sec:sourc}). To explore the interplay of non-linear sensor signals and the fringe tracking control loop, we designed a simplified model of the interferometer real-time control system, which is shown in Fig. \ref{fig:loopmodel}. 

The OPD and GD controllers are modeled by integrators, where the GD integrator is 20 times slower than the OPD integrator. The OPD sensor and actuator dynamics are modeled by a time delay of 3 ms. No noise sources are included and atmospheric dispersion is neglected. We assume the phase delay to be linear\footnote{In practice, the PD non-linearity amplitude is not zero and, in the laboratory, is about one order of magnitude smaller than the GD non-linearity.} whereas the non-linear group delay is modeled as
\begin{equation} 
GD = \frac{\lambda}{2 \pi} \phi  + a \sin( \nu \phi ), 
\end{equation}
where $a$ is the amplitude of the cyclic (sinusoidal) error with periodicity $\nu = \rho_\mathrm{cyclic}$ (Fig. \ref{fig:GDsimu}). $\nu$ is therefore measured in multiples of the carrier wavelength. For sinusoidal non-linearity, the relation between $a$ and $q$ is then 
\begin{equation}
q = \hat q \cos(\nu \phi ) \hspace{1cm} a = \hat q  \frac{\lambda}{2 \pi \nu },
\end{equation}
where $\hat q$ is the maximum value of $q$.

\begin{figure}[t]\begin{center} 
\includegraphics[width= 0.49\linewidth]{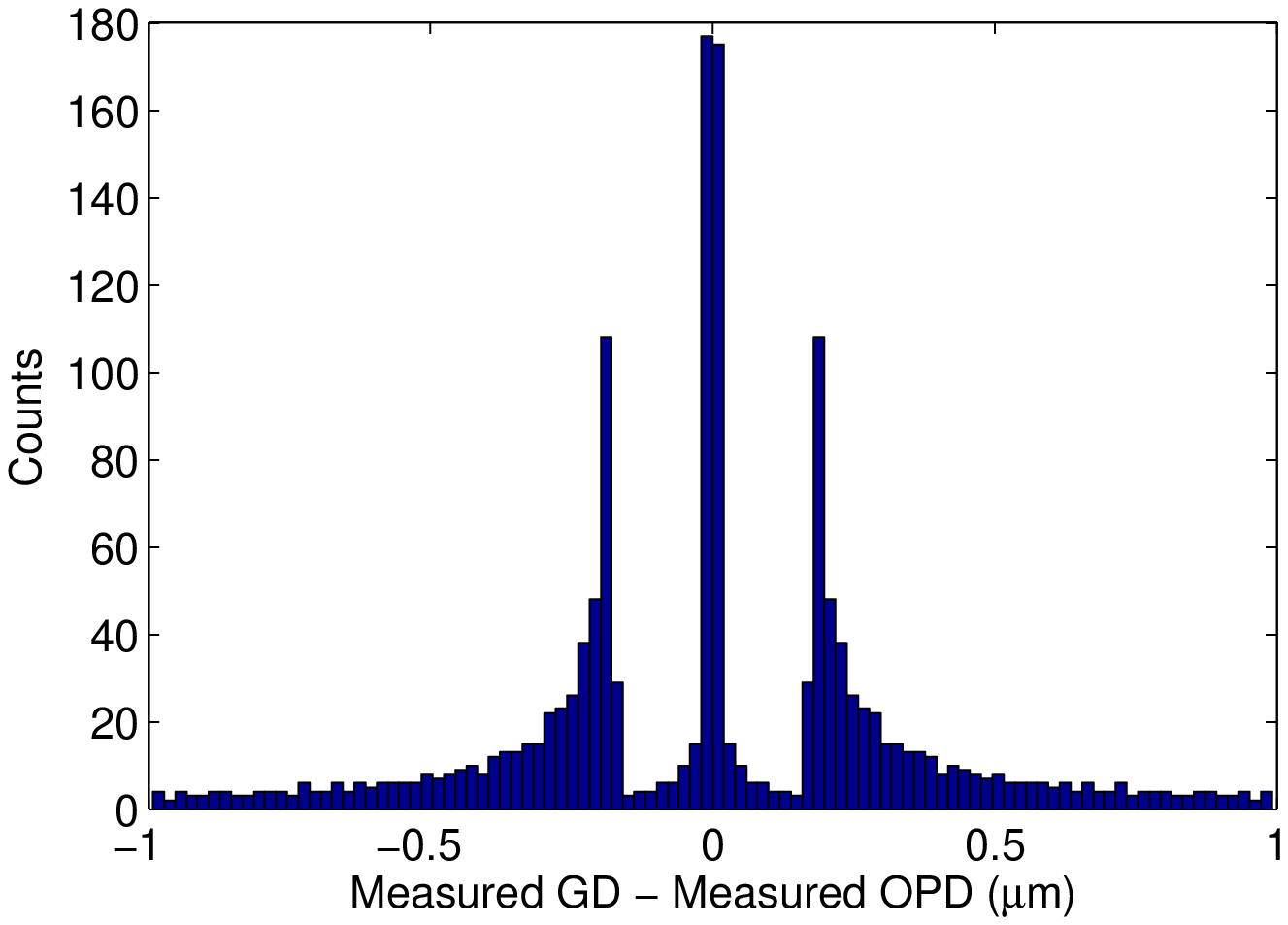}
\includegraphics[width= 0.49\linewidth]{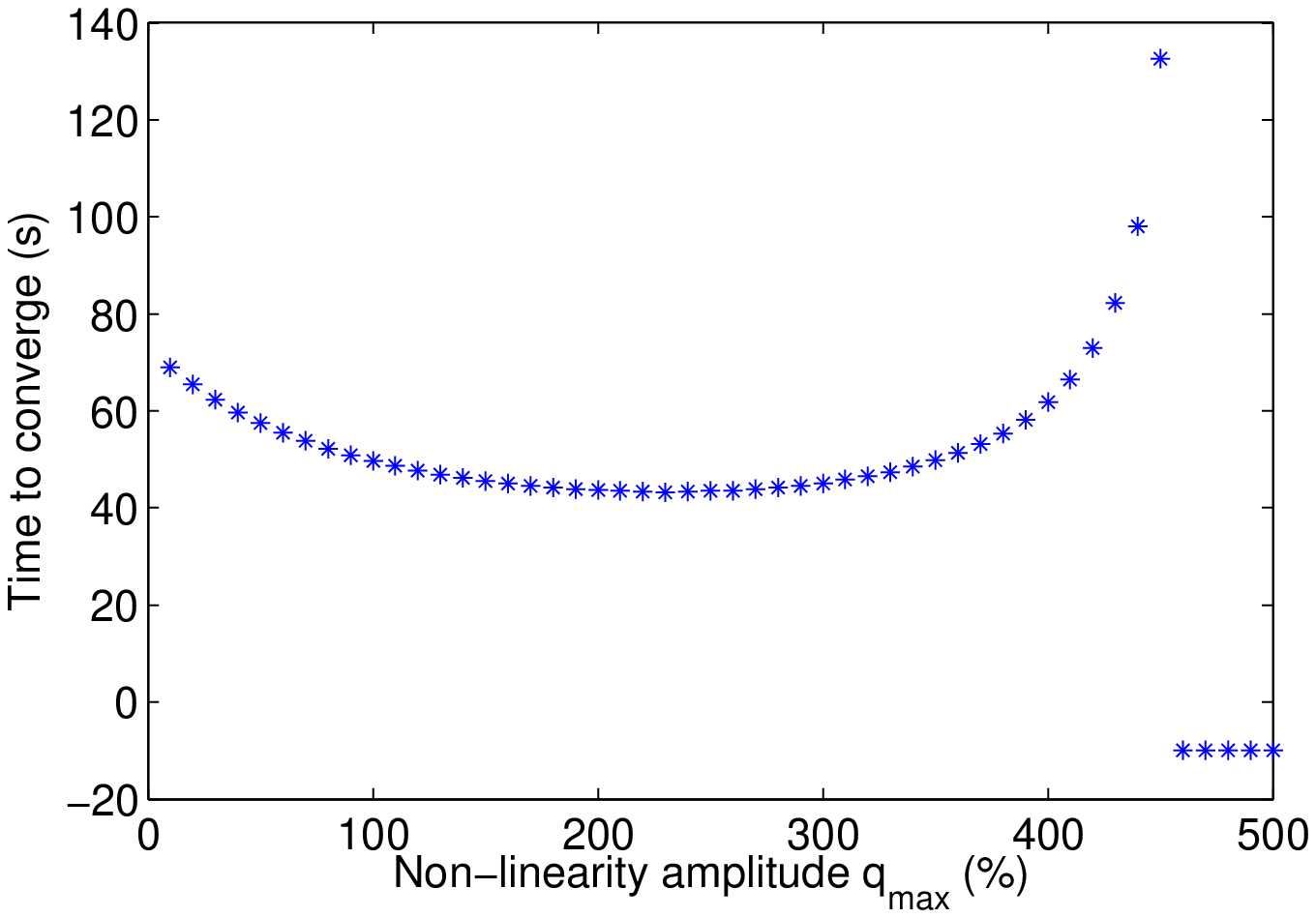} 
 \end{center} \caption{\emph{Left}: Simulated difference between measured GD and measured OPD for high non-linearity of $\hat q = 410$\,\%. \emph{Right}: Simulated convergence time as function of non-linearity amplitude $\hat q$. Negative values indicate situations where the system does not converge to \emph{true} zero GD, but the non-linearity is so high that several stable attractive points appear. The time-scales are not realistic in this simulation. In reality, the settling time of the fringe tracking loop amounts to a couple of seconds, once the fringes have been detected.}\label{fig:god}
\end{figure}

The result of this simulation is shown in Fig. \ref{fig:convergence}. When the measured group delay has a local maximum or minimum due to the non-linearity, the GD integrator slows down and produces nearly flat target OPD for the inner OPD loop for long times. Given the slow time constant, this behavior can be mistaken for a stable equilibrium point, which it is not. A good example is the dotted curve in Figure \ref{fig:convergence} between 10-40 s. The simulation can reproduce patterns similar to Figure \ref{fig:fringing}, as is shown in the histogram of Figure \ref{fig:god}. The difference between measured GD and measured OPD primarily adopts three equidistant values. The distance between these discrete levels depends on the non-linearity amplitude. An important characteristic of such a situation is that the measured group delay is close but not equal to zero. The closer the measured group delay is to zero at the local maximum/minimum, the longer the GD controller takes to get out of this meta-equilibrium. 

\begin{figure}\begin{center} 
\includegraphics[width= 0.5\linewidth]{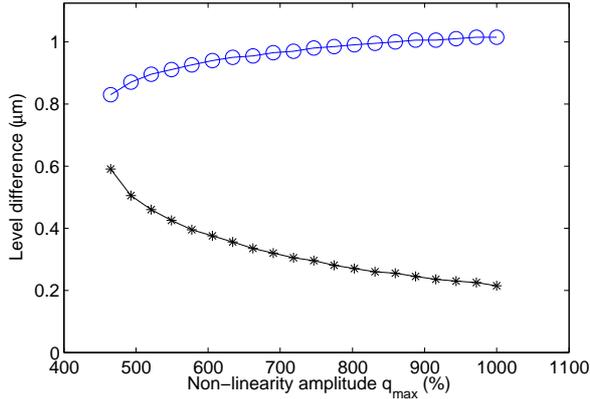} 
\end{center} 
\caption{Simulated difference between GD -- OPD levels for very high non-linearity amplitudes $\hat q$ that provoke several zero-crossings of GD and therefore several stable track points for the fringe tracking loop. The level difference is shown for cyclic frequencies of $\nu=1$ (black asterisks) and of $\nu=2$ (blue circles), assuming linear OPD measurements. The value of $\sim0.45\,\mu$m in Fig. \ref{fig:fringing2} is reproduced here if the dominant non-linearity has cyclic frequency of 1.}\label{fig:goddiff}
\end{figure}

If the non-linearity is so large that several zero-crossing occur, the control loop has several stable attractive points. The system has no way to distinguish between these attractive points and can remain locked on a false position (dash-dotted lines in Fig. \ref{fig:convergence}). In this case the measured GD is equal to zero. The difference between measured GD and measured OPD can have three equidistant values. The distance between these discrete levels depends on the non-linearity amplitude and is shown in Fig. \ref{fig:goddiff}. If this situation occurs in practice, the additional stable points may be less robust to noise, because the GD slope is larger in the central, desired stable point.

Another characteristic, that we can investigate using simulation is the settling time of the control loop, defined as the time it takes the system to reach the target GD value of 0. The simulated settling time is shown in the right panel of Figure \ref{fig:god}. It is found to be approximately stable for non-linearity amplitudes of $\hat q\sim 0-400$~\%. For even higher $\hat q$, the settling time increases rapidly as the local GD maxima and minima get very close to zero. If the group delay has several zeros, the systems locks on one of them and, since the simulation is noise-free, remains there. This is the case where the system measures a vanishing group delay but is far from the \emph{true} center of the fringe.

We have thus found two possible explanations for the the behaviour in Figures \ref{fig:fringing} and \ref{fig:fringing2}, which are both based on highly non-linear group delay measurements. Because the measured group delay is usually at zero (and not only close to it), the scenario in which several zero crossings are present appears more plausible. However, we do not have clear evidence and this guess is based on simulations which assume ideal and linear OPD measurements. 

\section{DISCUSSION AND CONCLUSIONS}
The FSU has demonstrated spatially-encoded fringe tracking at the VLTI. The first results are encouraging, especially in terms of the limiting magnitude and tracking residuals, which is essential for the PRIMA scientific programmes. The remaining important parameter is the fringe tracking accuracy, which appears to be ambiguous during FSU fringe tracking. The problem is difficult to detect and quantify because a second, independent fringe sensor is required to unambiguously measure the tracking accuracy.

Because the encountered problems can be caused by non-linear control signals, we have investigated the various sources of phase and group delay non-linearity. They are caused by errors that originate in the currently unavailable on-sky calibration of wavelengths and phase-shifts and in the inherent lack of real-time photometric monitoring capability of the FSU. The introduced perturbations equally affect the FSU estimates of visibility and signal-to-noise ratio, which is discussed in another contribution to these proceedings\cite{Schmid2010}.

The assessment of the impact of non-linear control signals required the investigation of the fringe-tracking control system architecture. The effect of GD non-linearity on the fringe-tracking loop behaviour was examined using a simplified model. Some patterns appearing during on-sky fringe tracking could be reproduced by the model. Although this is not a proof that the model describes the real situation, it gives a possible explanation for the observed behaviour. 

We conclude by listing the necessary steps to characterise and eventually improve the system:
\begin{itemize}
  \item Measure the FSU fringe-tracking accuracy with an independent fringe sensor attached to the system, which can be FINITO, the second FSU, MIDI, or AMBER. This will allow us to estimate the amplitude of the introduced biases. 
  \item Measure the non-linearity of FSU phase and group delay during on-sky observation. This requires dual-feed observations (first dual-feed fringe tracking is planned for July 2010). 
  \item Measure the phase-shifts and effective wavelengths during on-sky observation. In principle, this is possible by scanning the fringe packet fast enough to freeze the atmospheric turbulence, provided a precise delay measurement is available. Eventually, this will be part of the calibration plan for dual-feed astrometric observations and the procedure is described in\cite{Sahlmann:2009kx}. 
  \item Based on the \emph{qualitative} assessment of non-linearity sources presented here, a \emph{quantitative} study has to be undertaken to identify the factors that are currently limiting the performance. 
  \item A more sophisticated model of the control system that includes atmospheric noise, phase delay non-linearity and possibly atmospheric dispersion may be employed to study the fringe-tracking behaviour and its succeptibility to the various error sources in more detail.  
 \item A better model for the group-delay measurement principle has to be developed, in order to study the impact of calibration errors and to optimise the algorithm's robustness.
 \item Implement upgraded algorithms for group and phase delay, that account for different ABCD wavelengths.
\end{itemize}

As the fringe sensor and the data producing entity, the FSU is the central element of the PRIMA facility. The comprehensive characterisation of the system are vital for the scientific exploration of the data. This process has to be finalised before the scientific programme of extrasolar planet search with PRIMA astrometry\cite{Launhardt2008} can begin.



\acknowledgments     

J. S. would like to thank D. S\'egransan, J. B. Le Bouquin, and D.~Queloz for insightful discussions. The authors thank P.~Gitton for continuous support. The research of J. S. presented in this work has received funding from the European Community's Seventh Framework Programme under Grant Agreement 226604.
 

\bibliography{spie2010}   
\bibliographystyle{spiebib}   

\end{document}